\documentclass[12pt,a4paper,english]{article}
\usepackage[T1]{fontenc}
\usepackage[latin2]{inputenc}
\usepackage[english]{babel}
\usepackage{amsmath}
\usepackage{amsfonts}
\usepackage{indentfirst}
\usepackage[dvips]{graphicx}

\newcommand{\bea}{\begin{eqnarray}}
\newcommand{\eea}{\end{eqnarray}}
\newcommand{\be}{\begin{equation}}
\newcommand{\ee}{\end{equation}}

\newcommand{\Z}{{\mathbb Z}}

\newcommand{\C}{{\mathbb C}}

\def\G{\Gamma}

\newcommand{\cZ}{{\cal Z }} 
\newcommand{\Li}{{\rm Li}}

\begin{document}

\sloppy


\begin{flushright}
\begin{tabular}{l}
CALT-68-2812 \\

\\ [.3in]
\end{tabular}
\end{flushright}

\begin{center}
\Large{ \bf Refined matrix models from BPS counting}
\end{center}

\begin{center}

\bigskip

\bigskip 

Piotr Su{\l}kowski\footnote{On leave from University of Amsterdam and So{\l}tan Institute for Nuclear Studies, Poland.}

\bigskip

\bigskip

\medskip 

\emph{California Institute of Technology, Pasadena, CA 91125, USA} \\ [1mm]


\bigskip

\smallskip
 \vskip .4in \centerline{\bf Abstract}
\smallskip

\end{center}

We construct a free fermion and matrix model representation of refined BPS generating functions of D2 and D0 branes bound to a single D6 brane, in a class of toric manifolds without compact four-cycles. In appropriate limit we obtain a matrix model representation of refined topological string amplitudes. We consider a few explicit examples which include a matrix model for the refined resolved conifold, or equivalently five-dimensional $U(1)$ gauge theory, as well as a matrix representation of the refined MacMahon function. Matrix models which we construct have ordinary unitary measure, while their potentials are modified to incorporate the effect of the refinement.


\newpage 

\section{Introduction}

The purpose of this paper is to provide a free fermion and matrix model representation of refined topological string amplitudes, and more generally refined BPS counting functions in a system of D2 and D0 branes bound to a single D6-brane, in toric Calabi-Yau manifolds without compact four-cycles. Such a putative free fermion representation is interesting, as it would extend earlier results on wall-crossing in D6-D2-D0 system to the refined case. The motivation for finding a matrix model representation is as follows. In the non-refined case connections between such systems and matrix models are known from several perspectives. General relations between topological strings, gauge theories and matrix models were postulated by Dijkgraaf and Vafa \cite{DV}, and related to $\mathcal{N}=2$ theories in \cite{CachazoVafa}. Chern-Simons matrix model for the conifold and generalizations to lens spaces were considered in \cite{mm-lens,CSmatrix}. Explicit representation of partition functions of gauge theories and topological string theories on corresponding Calabi-Yau manifolds have been found in \cite{eynard-planch,SW-matrix,matrix2star}. Matrix model representation of partition functions on general toric manifolds has been found in \cite{EKM-I,EKM-II}. Matrix models encoding wall-crossing phenomena for a class of toric manifolds without compact four-cycles have been constructed in \cite{2010-OSY,SzaboTierz-DT,2010openWalls}. While partition functions of four-dimensional gauge theories can be encoded in hermitian matrix models, a generalization to five-dimensional theories, and more generally topological strings on toric manifolds, amounts to considering unitary matrix models \cite{CSmatrix,eynard-planch,SW-matrix,EKM-I,2010-OSY}. All these relations gained new interest with the formulation of the general matrix model solution in terms of the topological recursion \cite{eyn-or}, and the related remodeling conjecture postulated in the context of topological string theory \cite{BKMP}. One might therefore wonder if the relation between matrix models and topological strings, and more generally BPS counting, extends to the refined case as well. We also stress that worldsheet definition of the refined topological string theory is still not well understood, and the hope that matrix model reformulation might give some hint in this context is also an important motivation for this work.

Yet another motivation to study refinement from matrix model perspective arises from the AGT conjecture \cite{AGT}: as proposed in \cite{DV2009}, partition functions of four-dimensional, $\mathcal{N}=2$ theories can be encoded in so-called beta-deformed, hermitian matrix models. Certain aspects of this statement were tested in \cite{MMM-Selberg,MS-CIVDV,AY-beta,Yu-refine,MMS-DF,CDV-AGT,MMS-BGW,TowardsAGTmatrix,GeneralizedMatrixAGT,JWu,MMSbeta1}. In particular the appearance of the beta-deformed measure from the Nekarsov partition functions has been demonstrated also for both four- and five-dimensional gauge theories and certain topological string theories in \cite{2009betaMatrix}, however only to the leading order. On the other hand, the formalism of the topological recursion for hermitian models has been extended to the beta-deformed case \cite{BetaRecursion}. Therefore one might hope that the refined topological string theories could be encoded in unitary, beta-deformed matrix models. However, as explained and demonstrated explicitly in \cite{Klemm-Omega,Marino-beta}, this turns out not to be true even in the simple example of the resolved conifold. Nonetheless, due to deep consequences of the topological recursion \cite{eyn-or}, finding some matrix model representation of refined partition functions would be quite desirable; such matrix models would presumably arise as some deformation of a certain class of already known unitary matrix models. This is the task we cope with in this paper, not only from the viewpoint of topological string amplitudes, but also more generally in the context of BPS counting and wall-crossing phenomena. The refined matrix models which we find involve matrices of infinite size and have ordinary, unitary measure, while their potentials are modified in a way which encodes the refinement. We stress this is opposite to the beta-deformed models, whose measure is modified, however potentials are the same both in refined and non-refined cases. One immediate advantage of our result is the fact, that the topological recursion for models with undeformed measure \cite{eyn-or,BKMP} is much simpler and tractable than in the beta-deformed case \cite{BetaRecursion}, and could be readily applied to gain more insight into properties of refined amplitudes.

We recall that there are various definitions of refinement whose physical equivalence is not quite clear, however the agreement of the resulting exact solutions is a strong argument for an underlying common, general structure. In all these so-called refined theories a dependence on a single parameter, such as string coupling $g_s$ or the background $\hbar$ in gauge theories, is replaced by a dependence on two parameters, customarily denoted $\epsilon_1$ and $\epsilon_2$. In the context of gauge theory refined amplitudes arose from their formulation in the $\Omega$-background \cite{Nek}. In case of topological strings on non-compact, toric manifolds, refinement was introduced in terms of refined BPS counting, reformulated combinatorially in terms of the refined topological vertex \cite{refined-vertex,refined-Kanno}, and shown to agree with gauge theory results in $\Omega$ background in \cite{Taki,AwataKanno-refined}. From the viewpoint of AGT conjecture refined amplitudes are encoded in relevant conformal blocks of two-dimensional CFT, and the corresponding beta-deformed matrix models are characterized by the Vandermonde determinant raised to the power $\beta=-\epsilon_1/\epsilon_2$. In the context of wall-crossing and BPS counting in a system of D6-D2-D0 branes on toric manifolds, following and in parallel with non-refined developments in \cite{Szendroi,JaffMoore,ChuangJaff,YoungBryan,OY-crystals,WallM,2009pyramid,NagaoVO}, refined amplitudes were considered from physical and mathematical perspectives in \cite{RefMotQ,Nagao-open}. Among multitude chambers in which (refined) generating functions of D6-D2-D0 bound states are known, there is a special chamber in which they agree with topological string amplitudes on the same Calabi-Yau manifold, and in particular the agreement with the refined topological vertex calculation was shown in \cite{Nagao-open}. This is this last formulation of the refinement on which our derivation is based.

To find refined matrix models we follow a strategy which extends a non-refined presentation of \cite{2010-OSY}.\footnote{Our results were obtained independently and before an overlapping work \cite{LiuRefine1,LiuRefine2} appeared.} Firstly, generalizing the results of \cite{2009pyramid}, we construct free fermion representation of crystals representing refined BPS states in question. This allows to write the refined BPS generating functions $Z^{ref}_n$ in a chamber specified by $n$ as
\be
Z^{ref}_n = \langle \Omega^{ref}_+ | \overline{W}^{ref}_n |\Omega^{ref}_- \rangle,        \label{Zref-n}
\ee
where $|\Omega^{ref}_{\pm}\rangle$ are states representing a manifold in question, and $\overline{W}^{ref}_n$ are wall-crossing operators which determine a chamber of interest. Then we turn these fermionic correlators into a unitary matrix model form. The refined character of fermionic correlators results in a modified form of matrix model potentials. Similarly as in \cite{eynard-planch,SW-matrix,2010-OSY}, our potentials have nontrivial string coupling dependence to all orders. While our results are valid in all chambers, in the so called commutative chamber we obtain matrix model representation of refined topological string amplitudes. 

To briefly exemplify our results, we recall first that the refined topological string amplitude for the resolved conifold with K{\"a}hler parameter $Q$ (or equivalently five-dimensional, $U(1)$ gauge theory) is given by
\be   
\mathcal{Z}^{ref}_{top} = M(t_1,t_2) \prod_{k,l=0}^{\infty} (1 - Q t_1^{k+1} t_2^{l}),    \label{Zreftopconi}
\ee
where $t_1 = e^{-\epsilon_1}, t_2 = e^{\epsilon_2}$, and $M(t_1,t_2) = \prod_{k,l=0}^{\infty} (1- t_1^{k+1} t_2^l)^{-1}$ is the refined MacMahon function. To find a matrix model representation of $\mathcal{Z}^{ref}_{top}$, we first construct a general refined BPS generating function in the form (\ref{Zref-n}), where in the case of the conifold $n$ is a single integer. We then translate such a fermionic correlator into a matrix model form, and in $n\to\infty$ limit, which corresponds to the so-called commutative chamber, we find the matrix model representation (written in terms of eigenvalues $z_k=e^{i u_k}$) of the refined topological string amplitude
$$
\mathcal{Z}^{ref}_{top} = \int \mathcal{D} U \prod_k  \prod_{j=0}^{\infty}  \frac{ (1+z_k t_1^{j+1})\ (1+t_2^{j}/z_k)}{ (1+t_2^j Q/z_k)},
$$
where $\mathcal{D} U$  is the ordinary unitary measure (see (\ref{dU})). To the leading order the above integrand gives rise to the following potential
\be
V(u;\beta) =  \frac{1}{2}u^2 - (1-\beta^{-1})\textrm{Li}_2(-e^{i u}) - \textrm{Li}_2(-Q e^{- i u})  +  \mathcal{O}(g_s,\beta) .   \label{Vexample}
\ee 
In what follows we also present matrix models associated to other chambers of the K{\"a}hler moduli space. We can also immediately note that in the limit $Q\to 0$, the above result reduces to a matrix model representation of the refined MacMahon function, with the exact integrand given by a deformed theta-function, which in the genus expansion gives a $\beta$-deformation of the gaussian potential of the Chern-Simons matrix model (such that both the dilogarithm term, as well as $\mathcal{O}(g_s,\beta)$ corrections, vanish for $\beta=1$). In the main text we discuss in more detail other explicit results for $\C^3$, conifold, or resolution of $\C^3/\Z_2$ singularity. Similarly as in \cite{2010openWalls} we postulate a relation of those refined matrix integrands to open BPS amplitudes.

The paper is organized as follows. In section \ref{sec-ref} we recall definitions and basic properties of refined BPS invariants and introduce relevant notation. In section \ref{sec-fermions} we extend formalism of \cite{2009pyramid} to the refined case and present fermionic representation of refined generating functions. In section \ref{sec-matrix} we turn these refined fermionic results into matrix models and describe their properties. Section \ref{sec-discuss} contains a discussion.


\section{Refined wall-crossing in D6-D2-D0 system}   \label{sec-ref}

Refined degeneracies of D2 and D0-branes bound to a D6 brane on a Calabi-Yau manifold $X$ can be encoded in a generating function
$$
Z^{ref}_n(q,Q) = \sum_{\alpha,\gamma} \Omega^{ref}_{\alpha,\gamma}(n;y) q^{\alpha} Q^{\gamma},
$$
with D0-brane charge represented by $\alpha\in\Z$, D2-brane charge represented by $\gamma\in H_2(X,\Z)$, and a chamber in the K{\"a}hler moduli space specified by (possibly a set of parameters) $n$. Let $\mathcal{H}_{\alpha,\gamma}(n)$ denote a space of BPS states with given charges $\alpha,\gamma$ and asymptotic values of moduli corresponding to a chamber $n$, and $J_3$ denote a generator of the spatial rotation group. For fixed charges $\alpha,\gamma$ and a choice of chamber $n$, refined degeneracies
\be
\Omega^{ref}_{\alpha,\gamma}(n;y) = \textrm{Tr}_{\mathcal{H}_{\alpha,\gamma}(n)} (-y)^{2J_3},     \label{OmegaTr}
\ee
are interesting invariants if $X$ does not posses complex structure deformations, which is the case for non-compact, toric manifolds which we consider in this paper. These invariants were argued in \cite{RefMotQ} to agree with motivic Donaldson-Thomas invariants of \cite{KS}, and in the case of the resolved conifold the corresponding BPS generating functions were derived using the refined wall-crossing formula, and encoded in a refined crystal model. From mathematical viewpoint, and in terms of dimer models, such analysis was extended to quite a general class of toric manifolds without compact four-cycles in \cite{Nagao-open}, and shown therein to agree, in the commutative chamber, with refined topological vertex computations. For $y=1$ all these invariants reduce to ordinary non-refined invariants, whose generating functions were encoded in dimer or crystal models in \cite{ChuangJaff,YoungBryan,OY-crystals}, and represented in the free fermion formalism in \cite{2009pyramid,NagaoVO}. In the next section, based on definitions of BPS generating functions in terms of dimers or crystals constructed in \cite{RefMotQ,Nagao-open}, we will extend such free fermion formalism to the refined models.

Before proceeding we present in more detail a class of manifolds we are interested in. Similarly as in \cite{2009pyramid,2010-OSY}, we consider toric, non-compact Calabi-Yau manifolds without compact four-cycles, whose toric diagrams arise from a triangulation of a strip. Such a diagram consists of $N+1$ vertices, and there are $N$ $\mathbb{P}^1$'s in the geometry with K{\"a}hler parameters denoted $Q_p=e^{-T_p}$, $p=1,\ldots,N$. To each vertex in the diagram we associate a type $\tau_i=\pm 1$. If the local neighborhood of $\mathbb{P}^1$, represented by an interval between vertices $i$ and $i+1$, is $\mathcal{O}(-2)\oplus\mathcal{O}$, then $\tau_{i+1}=\tau_i$; if this neighborhood is of $\mathcal{O}(-1)\oplus\mathcal{O}(-1)$ type, then $\tau_{i+1}=-\tau_i$. We choose a type of the first vertex as  $\tau_1=+1$. 

We also need to introduce relevant notation for refined quantities. In the non-refined case the string coupling $g_s$ is related to the D0-brane charge as $q = e^{-g_s}$. The refinement is encoded in an additional parameter $\beta$. Instead $g_s$ and $\beta$ it is more convenient to use a pair of parameters
$$
\epsilon_1 = \sqrt{\beta} g_s, \quad \epsilon_2 = -\frac{g_s}{\sqrt{\beta}}, 
$$
so that $\beta = -\frac{\epsilon_1}{\epsilon_2},  \epsilon_1 \epsilon_2 = -g_s^2$. We often use the exponentiated counterparts
$$
t_1 = e^{-\epsilon_1}, \quad t_2 = e^{\epsilon_2},
$$
and also introduce 
$$
g_s B = \epsilon_1 + \epsilon_2 = g_s\big(\sqrt{\beta} - \frac{1}{\sqrt{\beta}} \big).
$$
The variable $y$ in (\ref{OmegaTr}) can be expressed as $y=t_1/q = q/t_2$, so that $y^2=t_1/t_2 = q^B$. In this notation the non-refined limit $y=1$ corresponds to $\beta=1$, for which $\epsilon_1=-\epsilon_2=g_s$ and $t_1=t_2=q$ and $B=0$.

 
With the above notation we can present some explicit BPS generating functions whose matrix model representation we are going to find. The simplest manifold one can consider is $\mathbb{C}^3$, for which one gets the refined MacMahon function \cite{refined-vertex}, see fig. \ref{fig-C3},
\be
Z^{\mathbb{C}^3} = M(t_1,t_2) = \prod_{k,l=0}^{\infty} \frac{1}{1 - t_1^{k+1} t_2^l}.   \label{Zbps-C3}
\ee
In this case there is no K{\"a}hler parameter, and therefore there are no interesting wall-crossing phenomena. 

We note that one could consider more general family of refinements parametrized by $\delta$, such that $M_{\delta}(t_1,t_2) = \prod_{k,l=0}^{\infty} \big(1 - t_1^{k+1+\frac{\delta-1}{2}} t_2^{l-\frac{\delta-1}{2}}\big)^{-1}$. For simplicity, in what follows we choose the value $\delta=1$ (note that in \cite{RefMotQ} another choice $\delta=0$ was made).

The resolved conifold provides a basic non-trivial example of wall-crossing, with a set of chambers parametrized by an integer $n$ (in the refined case one might also consider additional \emph{invisible} walls, which we do not discuss here). Corresponding refined generating functions were computed in \cite{RefMotQ} using a refined wall-crossing formula and in the chamber labeled by $n-1$ they read
\be
Z^{conifold}_{n-1} = M(t_1,t_2)^2 \Big( \prod_{k,l=0}^{\infty} \big(1 - Q t_1^{k+1} t_2^l\big) \Big) \Big( \prod_{k\geq 1,\ l\geq 0, \ k+l\geq n} \big(1 - Q^{-1} t_1^{k} t_2^l\big) \Big) \label{Zbps-coni}.
\ee
In the commutative chamber $n\to\infty$ the terms in the last bracket do not contribute anymore and the BPS generating function is simply related to the refined topological string amplitude given in (\ref{Zreftopconi})
$$
Z^{conifold}_{\infty} = M(t_1,t_2) \, \mathcal{Z}^{ref}_{top}.
$$
On the other hand, in the non-commutative chamber $n=0$, the refined generating function is given by the modulus square of the refined topological string amplitude.

For a resolution of $\mathbb{C}^3/\mathbb{Z}_2$ singularity there is also a discrete set of chambers parametrized by an integer $n$ and the corresponding BPS generating functions read
\be
Z^{\mathbb{C}^3/\mathbb{Z}_2}_{n-1} = M(t_1,t_2)^2 \Big( \prod_{k,l=0}^{\infty} \big(1 - Q t_1^{k+1} t_2^l\big)^{-1} \Big) \Big( \prod_{k\geq 1,\ l\geq 0, \ k+l\geq n} \big(1 - Q^{-1} t_1^{k} t_2^l\big)^{-1} \Big) .  \label{Zbps-C3Z2}
\ee

It is harder to write down generating functions for arbitrary chamber of an arbitrary geometry of our interest. However this can be done in for the non-commutative chamber of arbitrary geometry, where  -- similarly as in the non-refined case -- BPS generating function is given by the modulus square of the refined topological string amplitude
\be
Z^{ref}_0 = |\cZ^{ref}_{top}|^2 \equiv \cZ^{ref}_{top}(Q_i) \cZ^{ref}_{top}(Q_i^{-1}).     \label{Z-Ztop2}
\ee
The (instanton part of the) refined topological string amplitude is given by \cite{refined-vertex,Taki}
\be
\cZ^{ref}_{top}(Q_i) =  M(t_1,t_2)^{\frac{N+1}{2}} \prod_{k,l=0}^{\infty} \prod_{1\leq i < j \leq N+1} \Big(1- (Q_i Q_{i+1}\cdots Q_{j-1}) \, t_1^{k+1} t_2^l\Big)^{-\tau_i \tau_j},   \label{Ztop-strip}
\ee
with the notation introduced above.


\section{Refined wall-crossing and free fermions}     \label{sec-fermions}

The problem of counting bound states of D6-D2-D0 branes for local toric Calabi-Yau manifolds without compact four-cycles has been formulated in the free fermion formalism in \cite{2009pyramid,NagaoVO}. Among many advantages of such a representation is its immediate relation to melting crystals, as well as to matrix models, which was exploited in \cite{2010-OSY,2010openWalls}. Here we wish to extend such free fermion formalism to capture refined BPS invariants, as defined in \cite{RefMotQ,Nagao-open}. 

\begin{figure}[htb]
\begin{center}
\includegraphics[width=0.4\textwidth]{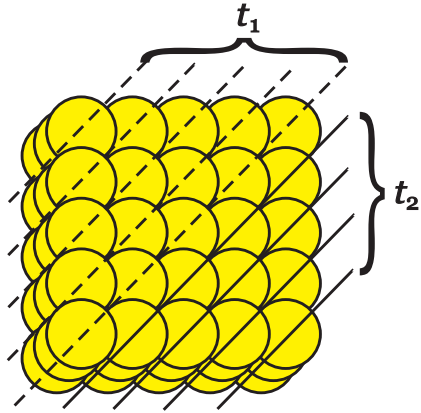} 
\begin{quote}
\caption{\emph{Refined plane partitions which count D6-D0 bound states in $\C^3$, as seen from the bottom (i.e. a negative direction of $z$-axis). Stones in each layer which intersects a dashed or solid line have weight $t_1$ or $t_2$ respectively. The resulting generating function is the refined MacMahon function $M(t_1,t_2)$. }} \label{fig-C3}
\end{quote}
\end{center}
\end{figure}

We consider first statistical models of colored pyramids. In the non-refined case \cite{2009pyramid}, to a geometry consisting of $N$ $\mathbb{P}^1$'s one associates a crystal which is sliced into layers in $N+1$ colors, denoted $q_0, q_1, q_2, \ldots, q_N$. 
In the non-refined case, parameters $q_1,\ldots, q_N$ encode K{\"a}hler parameters of the geometry $Q_1,\ldots,Q_N$, while the product $\prod_{i=0}^N q_i$ is mapped to (possibly inverse of) $q=e^{-g_s}$. In the refined case the assignment of colors is more subtle, as it must take into account a refinement of a single parameter $q$ into $t_1$ and $t_2$ introduced above. In particular, in the non-commutative chamber $q_{i\neq 0}$ are mapped (up to a sign, as in the non-refined case) to $Q_i$, however we will have to replace $q_0$ by two refined colors $q_0^{(1)}$ or $q_0^{(2)}$, so that $t_i = q_0^{(i)} q_1\cdots q_N$, for $i=1,2$. The simplest case of $\C^3$ refined plane partitions, discussed also in \cite{refined-vertex}, is shown in fig. \ref{fig-C3}. For other manifolds, in other chambers we will find more complicated assignment of colors. 

In \cite{2009pyramid} the structure and coloring of a given crystal, corresponding to a particular toric geometry, was encoded in fermionic states $|\Omega_{\pm}\rangle$, so that the generating function of BPS invariants could be written as a superposition of two such states (with additional insertion of wall-crossing operators in non-trivial chambers). In this section we construct a refined states $|\Omega^{ref}_{\pm}\rangle$ with similar properties. 
In the non-commutative chamber the states which we construct are such that
\be
Z^{ref}_0 = \langle \Omega^{ref}_+ | \Omega^{ref}_- \rangle.          \label{Z0ref}
\ee
We also construct refined version of wall-crossing operators $\overline{W}^{ref}_n$, such that the BPS generating function in $n$'th chamber can be written as
\be
Z^{ref}_n = \langle \Omega^{ref}_+ | \overline{W}^{ref}_n |\Omega^{ref}_- \rangle.    \label{Znref}
\ee

In section \ref{ssec-ncDT} below we construct states $| \Omega^{ref}_{\pm} \rangle$ for arbitrary manifold in a class of our interest. In section \ref{ssec-coni} we construct states $| \Omega^{ref}_{\pm} \rangle$ and wall-crossing operators $\overline{W}^{ref}_n$ for all chambers of the resolved conifold and a resolution of $\C^3/\Z_2$ singularity. We follow conventions used in \cite{2009pyramid,2010-OSY,2010openWalls}, which are summarized for convenience in appendix \ref{app-fermion}.


\subsection{Arbitrary geometry -- non-commutative chamber}   \label{ssec-ncDT}

\begin{figure}[htb]
\begin{center}
\includegraphics[width=\textwidth]{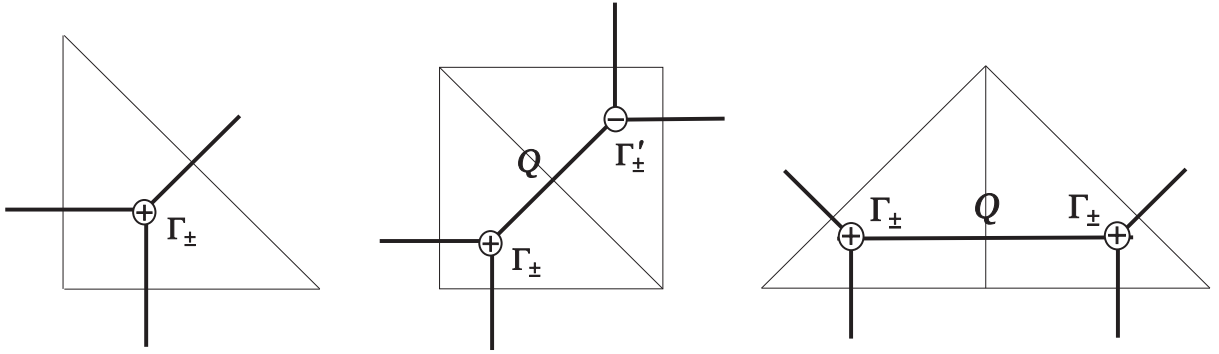} 
\begin{quote}
\caption{\emph{Toric diagrams and assignment of vertex operators in case of $\C^3$ (left), conifold (middle), and a resolution of $\C^3/\Z_2$ singularity (right). A sign $\oplus$ or $\ominus$ in each vertex denotes a corresponding type $\tau_i=\pm 1$. }} \label{fig-strips}
\end{quote}
\end{center}
\end{figure}

In this section we construct fermionic states $| \Omega^{ref}_{\pm} \rangle$, which allow to write the BPS generating functions in the non-commutative chamber as claimed in (\ref{Z0ref}). Similarly as in the non-refined case, the states $|\Omega^{ref}_{\pm}\rangle$ are constructed from an interlacing pattern of vertex $\G^{\tau_i}_{\pm}$ and weight operators. As the refinement does not modify the three-dimensional shape of the corresponding crystal, the assignment of vertex operators is the same as in the non-refined case \cite{2009pyramid} and can be similarly read off from the toric diagram. In particular, to the $i$'th vertex in the toric diagram (of type $\tau_i$ given above) we associate a vertex operator $\G_{\pm}^{\tau_i}(x)$, such that
$$
\G_{\pm}^{\tau_i=+1}(x)=\G_{\pm}(x),\qquad \qquad \G_{\pm}^{\tau_i=-1}(x)=\G_{\pm}'(x).
$$ 
Examples of this assignment for $\C^3$, conifold, and a resolution of$\C^3/\Z_2$ singularity are shown in fig. \ref{fig-strips}.

The structure which is modified in the refined case is the assignment of colors, which are encoded in the weight operators. A product of $N+1$ such operators $\G_{\pm}^{\tau_i}(x)$ is interlaced with weight operators in the following way. We introduce 
$N$ operators $\widehat{Q}_i$ representing colors $q_i$, for $i=1,\ldots,N$, and in addition two other colors $q_0^{(1)}$ and $q_0^{(2)}$, which are eigenvalues of $\widehat{Q}_0^{(1)}$ and $\widehat{Q}_0^{(2)}$. Operators $\widehat{Q}_1,\ldots,\widehat{Q}_N$ are associated to $\mathbb{P}^1$ in the toric diagram, and we define 
\be 
\widehat{Q}^{(i)} = \widehat{Q}_0^{(i)} \widehat{Q}_1\cdots \widehat{Q}_N,\qquad  t_{i} = q_0^{(i)} q_1 \cdots q_N, \qquad \textrm{for}\ i=1,2.    \label{t12}
\ee
Now we introduce
\bea
\overline{A}_{+}(x) & = & \G_{+}^{\tau_1} (x) \widehat{Q}_1 \G_{+}^{\tau_2} (x) \widehat{Q}_2 \cdots \G_{+}^{\tau_N} (x) \widehat{Q}_N \G_{+}^{\tau_{N+1}} (x) \widehat{Q}_0^{(1)}   ,       \nonumber        \\
\overline{A}_{-}(x) & = & \G_{-}^{\tau_1} (x) \widehat{Q}_1 \G_{-}^{\tau_2} (x) \widehat{Q}_2 \cdots \G_{-}^{\tau_N} (x) \widehat{Q}_N \G_{-}^{\tau_{N+1}} (x) \widehat{Q}_0^{(2)}.      \nonumber
\eea
Commuting all $\widehat{Q}_i$'s to the left or right we also introduce 
\bea
A_+(x) & = & \big(\widehat{Q}^{(1)}\big)^{-1} \, \overline{A}_{+}(x) = \G_{+}^{\tau_1} \big(xt_1\big)  \G_{+}^{\tau_2} \big(\frac{xt_1}{q_1}\big) \G_{+}^{\tau_3} \big(\frac{xt_1}{q_1 q_2}\big) \cdots \G_{+}^{\tau_{N+1}} \big(\frac{xt_1}{q_1q_2\cdots q_N}\big), \label{Aplus} \nonumber  \\
A_-(x) & = & \overline{A}_{-}(x) \, \big(\widehat{Q}^{(2)}\big)^{-1} = \G_{-}^{\tau_1} (x)  \G_{-}^{\tau_2} (xq_1) \G_{-}^{\tau_3} (x q_1 q_2) \cdots \G_{-}^{\tau_{N+1}} (x q_1q_2 \ldots q_N). \label{Aminus}   \nonumber
\eea
When the argument of any of these operators is $x=1$ we often use a simplified notation
$$
\overline{A}_{\pm} \equiv \overline{A}_{\pm}(1), \qquad A_{\pm} \equiv A_{\pm}(1). 
$$
Finally we can associate to a given toric manifold two states
\bea
\langle \Omega^{ref}_+| & = & \langle 0 | \ldots \overline{A}_+(1) \overline{A}_+(1) \overline{A}_+(1) = \langle 0 | \ldots A_+(t_1^2) A_+(t_1) A_+(1),  \label{Omega-plus}  \nonumber  \\
| \Omega^{ref}_- \rangle & = & \overline{A}_-(1) \overline{A}_-(1) \overline{A}_-(1) \ldots |0\rangle = A_-(1) A_-(t_2) A_-(t_2^2) \ldots |0\rangle ,  \label{Omega-minus}    \nonumber
\eea
where $|0\rangle$ is the fermionic Fock vacuum.

Our first claim is that the refined BPS generating function can be written as
\be
Z^{ref}_0 = \langle \Omega^{ref}_+ | \Omega^{ref}_- \rangle  \equiv \cZ_{top}(Q_i) \cZ_{top}(Q_i^{-1}),     \label{Z-cZ}
\ee
with $\cZ_{top}(Q_i)$ given in (\ref{Ztop-strip}), and under the following identification between $q_i$ parameters which enter a definition of $|\Omega^{ref}_{\pm}\rangle$ and string parameters $Q_i=e^{-T_i}$ (for $i=1,\ldots,N$)
\be
q_i = (\tau_i \tau_{i+1}) Q_i,   \label{qQ}    \nonumber
\ee
and with refined parameters $t_{1,2}$ identified as in (\ref{t12}). 
This result, in the special case of $\C^3$, conifold and $\C^3/\Z_2$ geometries, reproduces formulas (\ref{Zbps-C3}), (\ref{Zbps-coni}) and (\ref{Zbps-C3Z2}). 

To prove (\ref{Z-cZ}) for general geometry, we first note that commuting operators $A_+(x)$ with $A_-(y)$ 
$$
A_+(x) A_-(y) = A_-(y) A_+(x)\, C(x,y),
$$
gives rise to a factor 
$$
C(x,y) = \frac{1}{(1 - t_1 xy)^{N+1}} \prod_{1\leq i < j \leq N+1} \Big( \big(1 - (\tau_i \tau_j) x y t_1 (q_i q_{i+1}\ldots q_{j-1}) \big) \big(1 -  \frac{(\tau_i \tau_j) x y t_1}{ q_i q_{i+1}\ldots q_{j-1}} \big)  \Big)^{-\tau_i \tau_j}.
$$
Now we write the states $| \Omega^{ref}_{\pm} \rangle$ in terms of $A_{\pm}$ operators, and commute $\Gamma_{\pm}$ within each pair of $A_+$ and $A_-$ separately
$$
Z^{ref}_0 = \langle \Omega^{ref}_+ | \Omega^{ref}_- \rangle = \langle 0| \Big( \prod_{i=0}^{\infty} A_+(t_1^i) \Big) \Big( \prod_{j=0}^{\infty} A_-(t_2^j) \Big) |0\rangle = \prod_{i,j=0}^{\infty} C(t_1^i,t_2^j).
$$
This last product reproduces modulus square of the refined topological string partition function in (\ref{Z-cZ}) and therefore proves the claim (\ref{Z0ref}).


\subsection{Conifold and $\mathbb{C}^3/\mathbb{Z}_2$ -- all chambers}     \label{ssec-coni}

\begin{figure}[htb]
\begin{center}
\includegraphics[width=0.7\textwidth]{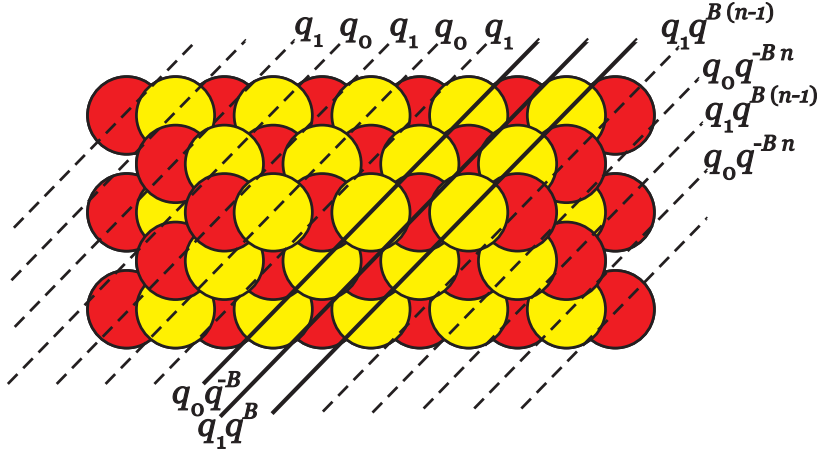} 
\begin{quote}
\caption{\emph{Refined pyramid crystal for the conifold, in the chamber corresponding to $n$ stones in the top row. Along each slice (as indicated by broken or solid lines) all stones have the same color, assigned as follows. On the left side (along broken lines), each light (yellow) and dark (red) slice has color denoted $q_0$ and $q_1$ respectively. Moving to the right, in the intermediate region (along solid lines), a color of each new light or dark slice is modified by respectively $q^{\mp B}$ factor (with respect to the previous light or dark slice). On the right side (again along broken lines), each light or dark slice has again the same color, respectively $q_0 q^{-B n}$ or $q_1 q^{B(n-1)}$. The assignment of colors in the intermediate region (along solid lines) interpolates between constant assignments on the left and right side of the pyramid.  }} \label{fig-colors}
\end{quote}
\end{center}
\end{figure}

\begin{figure}[htb]
\begin{center}
\includegraphics[width=0.4\textwidth]{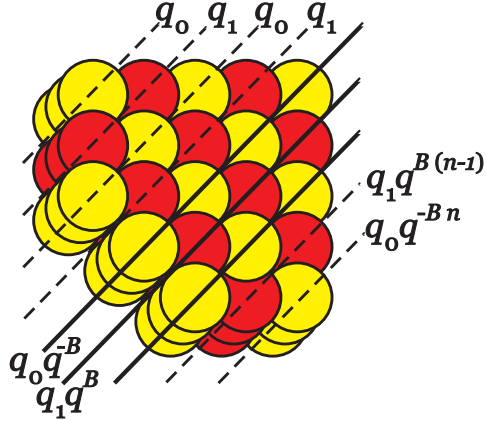} 
\begin{quote}
\caption{\emph{Refined pyramid crystal for the resolution of $\C^3/\Z_2$ singularity, in the chamber corresponding to $n$ stones in the top row, as seen from the bottom (i.e. a negative direction of $z$-axis). Even though the three-dimensional shape of the crystal is different than in the conifold case, the assignment of colors is the same, see fig. \ref{fig-colors}. }} \label{fig-C3Z2}
\end{quote}
\end{center}
\end{figure}

Fermionic representation can also be extended to non-trivial chambers. Even though this can be done for general geometry without compact four-cycles, for simplicity we restrict our considerations to the case of a conifold and a resolution of $\mathbb{C}^3/\mathbb{Z}_2$ singularity, which involve just one K{\"a}hler parameter $Q_1 \equiv Q$. In those cases, in a chamber labeled by $n-1$ we find the following representation of BPS generating function
\be
Z^{ref}_{n-1} = \langle \Omega^{ref}_+ | \overline{W}^{ref}_{n-1} |\Omega^{ref}_- \rangle,    \label{ZOmegaW}
\ee
where $\overline{W}^{ref}_{n-1}$ represents appropriate wall-crossing operator. In these both cases the toric diagram has two vertices, the first one of type $\tau_1=1$ and the second one denoted now $\tau\equiv \tau_2$, and $\tau=\mp 1$ respectively for the conifold and $\mathbb{C}^3/\mathbb{Z}_2$. A crystal associated to the expression (\ref{ZOmegaW}) has $n$ stones in the top row and can be sliced into interlacing single-colored layers. The assignment of colors is analogous as in the pyramid model discussed in \cite{RefMotQ,Nagao-open} (however our convention is slightly different, and corresponds to integer and non-symmetric, rather than half-integer and symmetric powers of $t_{1,2}$ in \cite{RefMotQ}). The pyramid crystal for the conifold is shown in  fig. \ref{fig-colors}. The coloring and weights for $\mathbb{C}^3/\mathbb{Z}_2$ are the same as for the conifold, even though the plane-partition shape of $\mathbb{C}^3/\mathbb{Z}_2$ crystal is different (though very analogous) than pyramid-like conifold crystal, see fig. \ref{fig-C3Z2}. Using the notation introduced above, the assignment of colors is determined as follows. 

All stones on one side of the crystal are encoded in the state 
$$
\langle \Omega^{ref}_+ | = \langle 0| \ldots \Big(\G_+(1) \widehat{Q}_1 \G_+^{\tau}(1) \widehat{Q}_0  \Big) \Big(\G_+(1) \widehat{Q}_1 \G_+^{\tau}(1) \widehat{Q}_0  \Big).
$$ 
The K{\"a}hler parameter $Q$, as well as the parameter $t_1$, are determined as
$$
q_1 = \tau Q t_1^{1-n},\qquad q_0 = \tau \frac{t_1^n}{Q},\qquad \textrm{so that} \quad q_0 q_1=t_1.
$$

Then the extended crystal, which has $n-1$ additional stones in the top row, is constructed by an insertion of the operator
\bea
\overline{W}^{ref}_{n-1} & = & \Big(\G_-(1) \widehat{Q}_1 \G_+^{\tau}(1) \widehat{Q}_0 \widehat{q^{-B}} \Big) \Big(\G_-(1) \widehat{Q}_1 \widehat{q^{B}} \G_+^{\tau}(1) \widehat{Q}_0 \widehat{q^{-2B}} \Big) \ldots \nonumber \\
& & \ldots \Big(\G_-(1) \widehat{Q}_1 \widehat{q^{(n-2)B}} \G_+^{\tau}(1) \widehat{Q}_0 \widehat{q^{(1-n)B}} \Big).  \nonumber
\eea
This operator consists of $n-1$ terms of the form $\Big(\G_-(1) \widehat{Q}_1 \widehat{q^{i B}} \G_+^{\tau}(1) \widehat{Q}_0 \widehat{q^{-(i+1)B}} \Big)$ for $i=0,\ldots,n-2$, where in each consecutive dark or light slice of stones we insert one additional operator $\widehat{q^{\pm B}}$, which changes the weight of each stone in this slice by $q^{\pm B} = (t_1/t_2)^{\pm 1}$ (with respect to the previous slice of the same light or dark color).

Finally, all stones on the right side of the crystal have again the same light or dark color, and the corresponding state reads
$$
|\Omega^{ref}_-\rangle =  \Big(\G_-(1) \widehat{Q}_1 \widehat{q^{(n-1)B}} \G_-^{\tau}(1) \widehat{Q}_0 \widehat{q^{-n B}} \Big) \Big(\G_-(1) \widehat{Q}_1 \widehat{q^{(n-1)B}} \G_-^{\tau}(1) \widehat{Q}_0 \widehat{q^{-n B}} \Big) \ldots |0\rangle.
$$
Therefore the varying weights in the middle range (along solid lines in fig. \ref{fig-colors} and \ref{fig-C3Z2}) interpolate between fixed weights of light and dark stones on two external sides of a crystal.

We can now commute away all weight operators in the above expressions, using relations from appendix \ref{app-fermion}. This leads to the representation
$$
Z^{ref}_{n-1} = \langle 0| \Big( \prod_{k=1}^{\infty} \G_+(t_1^k) \G_+^{\tau} (t_1^k/q_1) \Big)  \Big(\prod_{i=0}^{n-2} \G_-(t_2^i) \G_+^{\tau} (q_1^{-1} t_1^{-i}) \Big) \Big( \prod_{k=0}^{\infty} \G_-(t_2^{n-1+k}) \G_-^{\tau} (tQ t_2^k) \Big) |0\rangle  .
$$
Finally, commuting all vertex operators, we find
\be
Z^{ref}_{n-1} = M(t_1,t_2)^2 \, \prod_{k=1,l=0}^{\infty} (1 - Q t_1^{k} t_2^{l})^{-\tau} \, \prod_{k\geq 1,l\geq 0, k+l\geq n}^{\infty} (1 - Q^{-1} t_1^{k} t_2^{l})^{-\tau},
\ee
where $\tau=\mp 1$ respectively for the conifold and $\mathbb{C}^3/\mathbb{Z}_2$. This indeed reproduces (\ref{Zbps-coni}) and (\ref{Zbps-C3Z2}), and agrees (up to half-integer convention of $t_{1,2}$) with the results of \cite{RefMotQ}.


\section{Matrix models from fermions}     \label{sec-matrix}

Once the generating function of Donaldson-Thomas invariants is written in the fermionic formalism, it can be turned into matrix model form upon inserting appropriately chosen identity operator in the correlator (\ref{ZOmegaW})   
\be
Z^{ref}_{n-1} = \langle \Omega^{ref}_+ | \mathbb{I} \, \overline{W}^{ref}_{n-1} |\Omega^{ref}_- \rangle.       \label{ZOmega-I}
\ee
The identity operator $\mathbb{I}$ is represented by the complete set of states $|R\rangle\langle R|$ (representing two-dimensional partitions). We can use orthogonality relations of $U(\infty)$ characters $\chi_R$, and representation of these characters in terms of Schur functions $\chi_R=s_R(\vec{z})$ for $\vec{z}=(z_1,z_2,z_3,\ldots)$, to write
\bea
\mathbb{I} & = & \sum_R |R\rangle\langle R|  = \sum_{P,R} \delta_{P^t R^t} |P\rangle\langle R|  
=  \int \mathcal{D} U \sum_{P,R} s_{P^t}(\vec{z}) \overline{s_{R^t} (\vec{z})} |P\rangle\langle R| = \nonumber \\
& = & \int \mathcal{D}U \Big(\prod_k \G_-'(z_k)|0\rangle  \Big)  \Big(\langle0 | \prod_k \G_+'(z^{-1}_k) \Big),
\eea
where $\mathcal{D} U$ denotes the unitary measure written in terms of eigenvalues 
\be
\mathcal{D}U=\prod_k du_k \,\prod_{k<j}|z_k-z_j|^2,\qquad \qquad z_k=e^{iu_k}.    \label{dU}
\ee 

The identity operator in the above form can be inserted into (\ref{ZOmega-I}) which results in an expression involving only vertex operators $\G_{\pm}^{(\pm 1)}$. Then we can commute vertex operators away, again using relations from appendix \ref{app-fermion}, which leads to a matrix model with the unitary measure $\mathcal{D}U$. In the non-commutative chamber all factors arising from commuting these $\G_{\pm}^{(\pm 1)}$ operators depend on $z_k$ and contribute just to the matrix model potentials. In other chambers  additional factors may arise which are independent of $z_k$, and which, in a chamber labeled by $n$, contribute to some overall factor $f_n$. Thus in general we write the Donaldson-Thomas generating function as a matrix model in the form
\be
Z^{ref}_n = f_n\,\int \mathcal{D}U \prod_k e^{-\frac{\sqrt{\beta}}{g_s} V(z_k;\beta)},   \label{Zmatrix}
\ee
and it is convenient to introduce a factor $\sqrt{\beta}$ in front of the potential $V(z;\beta)$, or work with a rescaled coupling $g_s \beta^{-1/2}$.

In (\ref{ZOmega-I}) the identity operator has been inserted in a specific location. In fact there is large freedom of where this insertion should be chosen, which leads to various form of a matrix integrand. In \cite{2010openWalls} it has been shown that those various integrands can be identified with open BPS generating functions in various open chambers. We will also comment on a possible similar interpretation of refined integrands in what follows. However, let us first restrict to the a specific choice (\ref{ZOmega-I}) and discuss resulting matrix models. We use the following notation for a deformation of a theta-function 
\be
\Theta(z;t_1,t_2) = \prod_{j=0}^{\infty} (1 + z t_1^{j+1})(1+t_2^{j}/z)   \nonumber
\ee
to express certain integrands of matrix models that we come across.


\subsection{Arbitrary geometry -- non-commutative chamber}

As the first explicit example, we find matrix model representation of the refined BPS generating function in the non-commutative chamber. We start with the expression (\ref{ZOmega-I}) with no $\overline{W}^{ref}_{n-1}$ insertion, and use the form of $|\Omega^{ref}_{\pm}\rangle$ derived in section \ref{ssec-ncDT}. Performing the computation described above we get, in the non-commutative chamber for general geometry, the following matrix model:
\be
Z^{ref}_0 = \int \mathcal{D}U \prod_k \prod_{l=0}^N \Theta\big(\frac{\tau_{l+1} z_k }{q_1\cdots q_l} ; t_1,t_2\big)^{\tau_{l+1}} ,  \nonumber
\ee
i.e. we identify $e^{-\frac{\sqrt{\beta}}{g_s} V(z;\beta)} \equiv \prod_{l=0}^N \Theta\big(\tau_{l+1} z (q_1\cdots q_l)^{-1} ; t_1,t_2 \big)^{\tau_{l+1}}$.
The product over $l$ runs over all vertices and in this chamber we identify K{\"a}hler parameters $Q_p$ with weights $q_p$ via $q_p=(\tau_p \tau_{p+1})Q_p$.

Some special cases of the above result include:
\begin{itemize}
\item for $\mathbb{C}^3$ the generating function $Z^{ref}=M(t_1,t_2)$ is given by the refined MacMahon function (\ref{Zbps-C3}), and we find that the corresponding potential is a refined theta function
\be
e^{-\frac{\sqrt{\beta}}{g_s} V(z;\beta)}  =  \prod_{j=0}^{\infty} (1 + zt_1^{j+1})(1 + t_2^{j}/z) = \Theta(z;t_1,t_2)   \label{matrixC3}
\ee
\item for the conifold the non-commutative generating function $Z^{conifold}_0$ determined from (\ref{Zbps-coni}) gives rise to a matrix model with the following potential term  
$$
e^{-\frac{\sqrt{\beta}}{g_s} V(z;\beta)}  =  \prod_{j=0}^{\infty} \frac{(1 + z t_1^{j+1})(1 + t_2^{j}/z)}{(1 + z t_1^{j+1}/Q)(1 + \frac{Qt_2^{j}}{z})} = \frac{\Theta(z;t_1,t_2)}{\Theta(z/Q;t_1,t_2)}
$$
\item for $\mathbb{C}^3/\mathbb{Z}_{2}$ the non-commutative generating function $Z^{\C^3/\Z_2}_0$ determined from (\ref{Zbps-C3Z2}) gives rise to a matrix model with the following potential term  
\bea
e^{-\frac{\sqrt{\beta}}{g_s} V(z;\beta)} &  = & \prod_{j=0}^{\infty} (1 + z t_1^{j+1})(1 + t_2^{j}/z)  (1 + z t_1^{j+1}/Q)(1 + \frac{Qt_2^{j}}{z}) = \nonumber \\
& = & \Theta(z;t_1,t_2) \Theta(z/Q;t_1,t_2)  \nonumber 
\eea
\end{itemize}


\subsection{$\mathbb{C}^3$ matrix model}

Let us consider the simplest refined matrix model, corresponding to $\mathbb{C}^3$ geometry, with the exact potential given in (\ref{matrixC3}). Similarly as in \cite{eynard-planch,BEMS-Hurwitz}, one might expect that its behavior is governed by the leading order term in the potential.
Using the asymptotics 
$$
\log \prod_{i=1}^{\infty} \big(1 - z q^i\big) = -\frac{1}{g_s} \sum_{m=0}^{\infty} \Li_{2-m}\big(z\big) \frac{B_m g_s^m }{m!}
$$
this leading behavior reads
\be
e^{-\frac{\sqrt{\beta}}{g_s} V(z;\beta)}  = e^{-\frac{\sqrt{\beta}}{g_s} \big[ -\frac{1}{2}(\log z)^2 - (1-\beta^{-1})\textrm{Li}_2(-z)   + \mathcal{O}(g_s,\beta)  \big] }.    \label{C3-V0}
\ee
The first, quadratic term in the potential is the same as in the non-refined case. The term involving Li$_2(-z)$, as well as all higher order terms $\mathcal{O}(g_s,\beta)$, vanish for $\beta=1$. Therefore, for $\beta=1$, we obtain a Chern-Simons matrix model which indeed is known to give rise to MacMahon function in $N\to\infty$ limit \cite{CSmatrix,2010-OSY}. For general $\beta$, a resolvent $\omega(p)$ for a unitary model with the above potential can be found using the Migdal integral, as discussed in detail in \cite{2010-OSY,Marino}. This requires bringing the measure into a hermitian Vandermonde form, which introduces additional $T \log z$ term in the matrix potential, with the 't Hooft parameter $T=(g_s \beta^{-1/2}) N$. For the lowest order terms of the potential arising from (\ref{C3-V0}), this leads to
\be
\partial_z V(z;\beta) =  \frac{T - \log z - (1-\beta^{-1}) \log(z+1)}{z}.     \label{partialV}
\ee
Assuming a one-cut solution of the matrix model, and in terms of the rescaled coupling $g_s \beta^{-1/2}$, the Migdal resolvent is then given by\footnote{
A useful result \cite{2010-OSY} in such computations is $
\frac{1}{2T} \oint \frac{dz}{2\pi i} \frac{\log (z+c)}{z(p-z)}\frac{\sqrt{(p-a)(p-b)}}{\sqrt{(z-a)(z-b)}} =                  
 -\frac{1}{2 p T}\log \Big(\frac{\sqrt{(a+c)(b-p)} - \sqrt{(b+c)(a-p)}}{(p+c)(\sqrt{b-p}-\sqrt{a-p})}   \Big)^2       
 -\frac{\sqrt{(p-a)(p-b)}}{2 p T \sqrt{a b}} \log\Big( \frac{\sqrt{(a+c)b}-\sqrt{(b+c)a}}{c(\sqrt{a}-\sqrt{b})}  \Big)^2$.
This arises from contour integrals around poles $z=0$ and $z=p$, as well as along the branch cut of the logarithm $(-\infty,-c)$ which is found using $\int\frac{dx}{(x-p)\sqrt{(x-a)(x-b)}} = -\frac{1}{\sqrt{(p-a)(p-b)}} \log \frac{(\sqrt{(x-a)(b-p)} - \sqrt{(x-b)(a-p)})^2}{(p-x)\sqrt{(p-a)(p-b)}}.$
} 
\be
\omega(p) = \frac{1}{2 T} \oint \frac{dz}{2\pi i} \frac{\partial_z V(z)}{p-z}\frac{\sqrt{(p-a_-)(p-a_+)}}{\sqrt{(z-a_-)(z-a_+)}},   \nonumber   \label{Migdal}
\ee
so that the endpoints of the cut $a_-$ and $a_+$ are encircled counter-clockwise by the integration contour. Moreover, one has to impose the following consistency condition on the resolvent 
\be
\lim_{p\to\infty} \omega(p) = \frac{1}{p},  \nonumber     \label{Migdal-norm}    
\ee 
which imposes certain conditions on the end-points of the cut $a_{\pm}$. We find that for the potential (\ref{partialV}) these conditions take form
\bea
\frac{2}{\sqrt{a_-} + \sqrt{a_+}}\Big( \frac{2}{\sqrt{a_-+1} + \sqrt{a_+ +1}} \Big)^{(1-\beta^{-1})} & = & e^{T/2}, \label{ab-eqns1} \\   
\frac{\sqrt{a_-} + \sqrt{a_+}}{\sqrt{a_- a_+}}\Big( \frac{\sqrt{a_-} + \sqrt{a_+}}{\sqrt{(a_- +1)a_+} + \sqrt{(a_+ +1)a_-}} \Big)^{(1-\beta^{-1})} & = & 2e^{-T/2}.   \label{ab-eqns2}
\eea
For the non-refined case $\beta=1$ these equations simplify and can be exactly solved \cite{2010-OSY}. For arbitrary $\beta$ the cut end-points found in \cite{2010-OSY} get corrections in $(1-\beta^{-1})$,
$$
a_{\pm} = -1  + 2e^{-T} \pm  2 i e^{-T/2} \sqrt{1 - e^{-T}}  +  \mathcal{O}(1-\beta^{-1}),
$$
which leads to a $\beta$-deformed spectral curve. To find these corrections $\mathcal{O}(1-\beta^{-1})$ in the exact form appears not easy, and it would be interesting to compare the resulting curve with the quantum curve of the beta-deformed formalism of \cite{BetaRecursion}. In particular they both give rise to the same result in the four-dimensional limit \cite{Marino-beta}, so understanding a discrepancy of the five-dimensional results is an important issue. It would also be interesting to find the partition function for the above model with finite 't Hooft coupling $T$, and verify if it is related to the refined conifold topological string amplitude, as is indeed the case in the non-refined case.

As already mentioned before, we can also obtain more general matrix models by inserting the identity operator in various places in the fermionic representation of BPS function. In particular, inserting it at position $k$ in a string $\overline{A}_-$ operators in (\ref{Omega-plus}) in $\C^3$ case, we get the following representation
\be
Z^{ref}_k = f_k \int \mathcal{D}U \,  \prod_{j=0}^{\infty} (1 + z t_1^{j+1})(1 + t_1^k t_2^{j}/z),   \label{matrixC3-k}   \nonumber
\ee
with the prefactor $f_k=M(t_1,t_2) \prod_{j=k}^{\infty} \prod_{i=0}^{\infty} (1 - t_1^j t_2^i)$. In the non-refined case in \cite{2010openWalls}, the above integrand with identification $t_1=t_2=q$ was related to open BPS generating function in an open chamber labeled by $k$. It would be interesting to extend such an interpretation to the refined case too. In particular, we note that in the limit $k\to\infty$, which should correspond to the ordinary open topological string amplitude, the above integrand indeed reduces to one particular form of a refined brane partition function in $\C^3$ computed in \cite{refined-vertex}.

\subsection{Conifold -- all chambers}

Using the representation (\ref{ZOmega-I}) and fermionic results found in section \ref{ssec-coni}, we find the following matrix model for the conifold in the $n$'th closed BPS chamber (corresponding to a pyramid with $(n+1)$ stones on top), and with identity representation (\ref{identity}) inserted at position $k$ in a string $\overline{A}_-$ operators in (\ref{Omega-plus})
\be
Z^{ref}_{n,k} = f_{n,k}(q,Q)  \int \mathcal{D}U \prod_l \prod_{j=0}^{\infty}  \frac{ (1+z_l t_1^{j+1})\ (1+t_2^{j} t_1^k /z_l)}{ (1+z_l t_1^{j+n+1}/Q)\ (1+t_2^j Q t^k/z_l)}     \label{Zref-nk-coni}
\ee
with relevant prefactor $f_{n,k}(q,Q) $. In the special case $k=0$ we get
\bea
Z^{ref}_{n} & = & M(t_1,t_2)^2 \, \prod_{k=1,l=0}^{\infty} (1 - Q t_1^{k} t_2^{l}) \, \prod_{k\geq 1,l\geq 0, k+l\geq n+1}^{\infty} (1 - Q^{-1} t_1^{k} t_2^{l}) = \label{MM-Zn-conifold}    \nonumber  \\
& = & f_{n}(q,Q) \int \mathcal{D}U \prod_k \prod_{j=0}^{\infty}  \frac{ (1+z_k t_1^{j+1})\ (1+t_2^{j}/z_k)}{ (1+z_k t_1^{j+n+1}/Q)\ (1+t_2^j Q/z_k)},    \nonumber
\eea
with the prefactor given by
\be
f_{n}(q,Q) = \Big( \prod_{i=1}^{n} \prod_{k=0}^{\infty} \frac{1}{1-t_1^{i} t_2^k} \Big) \Big( \prod_{i=1}^{n} \prod_{j=n+1-i}^{\infty} (1-t_1^{i} t_2^j/Q) \Big).    \nonumber    \label{fn-conifold}
\ee
We determine cut end-points, which would follow from deformed equations analogous to (\ref{aAll}) and (\ref{bAll}), and which would lead to a $\beta$-deformed mirror curve; however this is quite involved technically. 

On the other hand, it is instructive to analyze the limit of commutative chamber $n\to \infty$ . Firstly, in this limit the prefactor simplifies to $f_{\infty} = M(t_1,t_2)$. We also know that the total value of the BPS generating function reduces in this limit to the refined topological string amplitude, up to a single factor of $M(t_1,t_2)$. In consequence in the commutative chamber we get a matrix model representation of the refined topological string amplitude for the conifold
\bea
\mathcal{Z}^{ref}_{top} & = & M(t_1,t_2) \prod_{k,l=0}^{\infty} (1 - Q t_1^{k+1} t_2^{l}) = \nonumber \\
& = & \int \mathcal{D}U \prod_k  \prod_{j=0}^{\infty}  \frac{ (1+z_k t_1^{j+1})\ (1+t_2^{j}/z_k)}{ (1+t_2^j Q/z_k)}.    \nonumber
\eea
Now the leading order potential term is a modification of the $\C^3$ potential (\ref{C3-V0}) by a $Q$-dependent dilogarithm term
\be
V(z;\beta) = -\frac{1}{2}(\log z)^2 - (1-\beta^{-1})\textrm{Li}_2(-z) - \textrm{Li}_2(-Q/z) + \mathcal{O}(g_s,\beta). \label{VconiTop}
\ee
In the limit $Q\to 0$ the above topological string partition function becomes just the refined MacMahon function, and the matrix integral consistently reproduces  $\C^3$ result (\ref{matrixC3}).


\subsection{$\mathbb{C}^3/\mathbb{Z}_2$ -- all chambers}

The results for $\mathbb{C}^3/\mathbb{Z}_2$ arise similarly as those for the conifold. Using the representation (\ref{ZOmega-I}) and fermionic construction from section \ref{ssec-coni} we find
\bea
Z_n & = & M(t_1,t_2)^2 \, \prod_{k=1,l=0}^{\infty} (1 - Q t_1^{k} t_2^{l})^{-1} \, \prod_{k\geq 1,l\geq 0, k+l\geq n}^{\infty} (1 - Q^{-1} t_1^{k} t_2^{l})^{-1} = \label{MM-Zn-C3Z2}   \nonumber \\
& = & f_n(q,Q) \int \mathcal{D}U \prod_k \prod_{j=0}^{\infty}  (1+z_k t_1^{j+1})\ (1+t_2^{j}/z_k) (1+z_k t_1^{j+n+1}/Q)\ (1+t_2^j Q/z_k),    \nonumber
\eea
where
\be
f_n(q,Q) = \Big( \prod_{i=1}^{n} \prod_{k=0}^{\infty} \frac{1}{1-t_1^{i} t_2^k} \Big) \Big( \prod_{i=1}^{n} \prod_{j=n+1-i}^{\infty} \frac{1}{1-t_1^{i} t_2^j/Q} \Big).      \nonumber   \label{fn-C3Z2}
\ee 
In particular, in the commutative chamber $n\to \infty$ we get again $f_{\infty} = M(t_1,t_2)$. Therefore in the commutative chamber we get a matrix model representation of the refined topological string amplitude
\be
\mathcal{Z}^{ref}_{top} = M(t_1,t_2) \prod_{k=1,l=0}^{\infty} \frac{1}{1 - Q t_1^{k} t_2^{l}} = \int \mathcal{D}U \prod_k  \prod_{j=0}^{\infty}  (1+z_k t_1^{j+1}) (1+t_2^{j}/z_k) (1+t_2^j Q/z_k) .  \nonumber
\ee
In the limit $Q\to 0$ we again recover the refined MacMahon function, as well as the expected integrand of $\C^3$ matrix model (\ref{matrixC3}). In this case it is also straightforward to find more general matrix models, analogous to (\ref{Zref-nk-coni}), which would presumably be related to refined open amplitudes.


\section{Discussion}    \label{sec-discuss}

In this paper we have found a free fermion, as well as a unitary matrix model representation of refined BPS generating functions of D0 and D2-branes bound to a single D6-brane, and in particular topological string amplitudes, in toric Calabi-Yau manifolds without compact four cycles. We mainly considered explicit examples of $\C^3$, conifold, and $\C^3/\Z_2$ geometries, as well as an arbitrary geometry in the non-commutative chamber, however generalization to other chambers for manifolds in this class is straightforward. A general consequence of our results is the fact that refined generating functions, at least for the class of manifolds which we considered, have nice properties of ordinary matrix model expressions \cite{eyn-or,BKMP}, such as integrability, symplectic invariance of associated free energy coefficients $F_g$, automatic appearance of the whole family of differentials $W^g_n$, etc. One advantage of our representation is that these properties are much better understood for ordinary matrix models, rather than for matrix models for beta-ensembles \cite{BetaRecursion}, which in fact are known \emph{not} to reproduce the refined topological string amplitudes \cite{Klemm-Omega,Marino-beta}. It is also important to understand a difference between these two beta-deformations. As follows from the results of \cite{Marino-beta}, in case of the conifold (or five-dimensional $U(1)$ gauge theory), the four-dimensional gauge theory limits of both deformations agree. Understanding the origin of a discrepancy in five-dimensional deformation should lead to interesting new insights.

There are many other questions which require further investigation. Firstly, a nontrivial task is to find spectral curves of our models. As we discussed, these would be $\beta$-deformation of curves found in a non-refined case in \cite{2010-OSY}. Having known such curves would allow to apply the topological recursion to recover quantities $W^g_n$ and $F_g$ explicitly from matrix model perspective. This appears nontrivial, in particular due to all order $g_s$ corrections to our potentials. However these corrections arise from terms involving quantum dilogarithms. Potentials which involve quantum dilogarithms were considered also in \cite{eynard-planch,BEMS-Hurwitz}, where it was shown that higher $g_s$ essentially do not modify resulting invariants, and one can effectively consider a leading order contribution to the potential, similarly as in (\ref{C3-V0}) and (\ref{VconiTop}) in our case. It would be interesting to confirm if analogous phenomenon takes place for the potentials which we consider.

Furthermore, it would be interesting to extend our discussion to the open string case, on one hand refining the discussion in \cite{2010openWalls} and providing M-theory derivation of putative open BPS generating functions, and on the other relating $W^g_n$ to brane amplitudes in matrix models in the topological string limit. In particular this should provide a deeper understanding of nontrivial prefactors in intermediate chambers.

It would of course be interesting to extend our results to toric manifolds with compact four-cycles, in particular those related by geometric engineering to gauge theories. This might be possible by considering more involved crystal models, such as those in \cite{QuiversCrystals}.

Among other questions, it is interesting what our matrix models compute for finite size of matrices $N$. It was shown in \cite{2010-OSY} that in the non-refined case finite $N$ engineers more complicated toric manifolds with an additional two-cycle (as is already the case in the Chern-Simons matrix models \cite{CSmatrix}, where a finite 't Hooft coupling encodes the size of the single $\mathbb{P}^1$ of the resolved conifold). In particular it is tempting to speculate whether the matrix model (\ref{matrixC3}) with finite $N$ would also provide the refined conifold topological string partition function.

It would also be interesting to understand the issues of holomorphic anomaly and modularity and make contact with discussions in \cite{Klemm-Omega,matrix-anomaly,KreflWalcher-extended}, and more generally with large literature on refined invariants. 

We hope that continuing this line of research would be a rewarding experience.


\bigskip

\bigskip

\begin{center}
{\bf Acknowledgments}
\end{center}

\medskip

I am grateful to Hirosi Ooguri for inspiring discussions and comments on the manuscript. I thank Andrea Brini, Yu Nakayama and Jaewon Song for useful, refined conversations. I appreciate kind hospitality of the Simons Workshop in Mathematics and Physics (2010) where parts of this work were done. This research was supported by the DOE grant DE-FG03-92ER40701FG-02 and the European Commission under the Marie-Curie International Outgoing Fellowship Programme. The contents of this publication reflect only the views of the author and not the views of the funding agencies. 



\newpage 

\appendix

\section{Free fermion formalism}   \label{app-fermion}

In this appendix we summarize free fermion formalism used in \cite{2009pyramid,2010-OSY,2010openWalls}. We consider the Heisenberg algebra
$[\alpha_m,\alpha_{-n}] = n \delta_{m,n}$ and define vertex operators
$$
\G_{\pm}(x) = e^{\sum_{n>0} \frac{x^n}{n}\alpha_{\pm n}}, \qquad \qquad \G'_{\pm}(x) = e^{\sum_{n>0} \frac{(-1)^{n-1}x^n}{n}\alpha_{\pm n}},
$$
which satisfy commutation relations
$$
\G_+(x) \G_-(y) = \frac{1}{1-xy} \G_-(y) \G_+(x),   \qquad \quad
\G'_+(x) \G'_-(y) = \frac{1}{1-xy} \G'_-(y) \G'_+(x),   
$$
$$
\G'_+(x) \G_-(y) = (1+xy) \G_-(y) \G'_+(x),   \qquad \quad
\G_+(x) \G'_-(y) = (1+xy) \G'_-(y) \G_+(x).
$$
These operators act on fermionic states $|\mu\rangle$, corresponding to two-dimensional partitions $\mu$, as 
\bea
\G_-(x) |\mu\rangle  =  \sum_{\lambda \succ \mu} x^{|\lambda|-|\mu|}|\lambda\rangle, & & \qquad \qquad  
\G_+(x) |\mu\rangle  =  \sum_{\lambda \prec \mu} x^{|\mu|-|\lambda|}|\lambda\rangle,      \\
\G'_-(x) |\mu\rangle =  \sum_{\lambda^t \succ \mu^t} x^{|\lambda|-|\mu|}|\lambda\rangle,& & \qquad \qquad
\G'_+(x) |\mu\rangle  =  \sum_{\lambda^t \prec \mu^t} x^{|\mu|-|\lambda|}|\lambda\rangle,
\eea
where $\prec$ is the interlacing relation. We also consider various weight operators $\widehat{Q}_g$, with eigenvalues representing colors and denoted $q_g$, such that
$$
\widehat{Q}_g|\lambda\rangle = q_g^{|\lambda|}|\lambda\rangle,
$$
and their commutation relations with vertex operators read
\bea
\G_+(x) \widehat{Q}_g = \widehat{Q}_g \G_+(x q_g), & & \qquad \qquad
\G'_+(x) \widehat{Q}_g = \widehat{Q}_g \G'_+(x q_g), \\
\widehat{Q}_g \G_-(x) = \G_-(x q_g) \widehat{Q}_g, & & \qquad \qquad 
\widehat{Q}_g \G'_-(x) = \G'_-(x q_g) \widehat{Q}_g.
\eea



\end{document}